\documentstyle{elsart} 

\input psfig
\input epsf

\def\xslide#1#2#3#4#5#6{\centerline{\psfig
{figure=#1,height=#2,bbllx=#3bp,bblly=#4bp,bburx=#5bp,bbury=#6bp,clip=}}}

\def\d3p{d^3p}
\def\tp3{(2\pi)^3}

\begin{document}

\hfill INP 1728/PH
\vspace{5mm}

\begin{frontmatter}
\title{Melting of the quark condensate in the NJL model with meson loops} 
\thanks{Research supported in part by the Polish State Committee for
        Scientific Research, grant 2P03B~188~09, by the
        Stiftung f\"ur Deutsch-Polnische Zusammenarbeit, project 
        1522/94/LN, and by the Maria Sk\l{}odowska-Curie grant PAA/NSF-94-158}
\author{Wojciech Florkowski and Wojciech Broniowski}
\address{H. Niewodnicza\'nski Institute of Nuclear Physics,
         PL-31342 Krak\'ow, Poland}

\begin{abstract}
Temperature dependence of the quark condensate is 
studied in the Nambu--Jona-Lasinio model 
with meson loops. Substantial differences are found 
compared to the results with quark loops only.
\end{abstract}
\end{frontmatter}

\section{Introduction}
\label{sec:introduction}

It is generally believed that QCD undergoes chiral restoration
at sufficiently high temperatures. This is supported by lattice
simulations \cite{K95}, as well as by a variety of model 
calculations. As the temperature grows, the value
of the quark condensate increases from its negative $T=0$ value 
and approaches zero. As shown in Ref. \cite{GL}, in the exact chiral 
limit (zero current quark masses) chiral symmetry dictates the form of the
first two terms ($\sim T^2$ and $T^4$) in the low-temperature expansion 
of the quark condensate. At higher temperatures we do not have
fundamental knowledge of the behavior of $\langle \overline{q}q \rangle$, 
however most of model calculations show a phase transition at
temperatures $T \sim {\rm 150 - 200}$~MeV. Lattice calculations
also show a dramatic change of $\langle \overline{q}q \rangle$ 
at similar temperatures. 

In this letter we study the temperature dependence of the quark condensate 
in the two-flavor Nambu--Jona-Lasinio (NJL) model \cite{NJL}. There have 
already been several studies \cite{HK85,BMZ87,Lutz92} of chiral restoration 
in this model. Our investigation brings a new important element: it includes 
{\em meson loops} in a self-consistent way. 
Previous studies have been performed at the quark-loop level only. 
Attempts have been made to include 
meson loops in the NJL model, but self-consistency was not 
completely 
fulfilled \cite{Heid}. 
Our approximation is symmetry-conserving \cite{DSTL95,NBCRG96}, 
hence it is consistent 
with all requirements of chiral symmetry.
The key ingredient is the self-consistency 
in solving the equation for the scalar density with meson loops present.
This makes the approach consistent with the requirements of 
chiral symmetry, such as the Goldstone theorem, Gell-Mann--Oaks--Renner
and Goldberger-Treiman relations, or one-loop chiral expansions.

We find important qualitative and quantitative differences in
the temperature dependence of the quark condensate 
in our calculation with meson loops 
compared to the case with quark loops only.
With quark loops only, at low temperatures the condensate
remains flat, whereas in our case it changes considerably. We show that  
in the exact chiral limit the change agrees with the prediction
of the chiral perturbation theory \cite{GL}. We also find that 
meson loops decrease the temperature of chiral restoration by about 10\%.

\section{Definition of the model}
\label{sec:model}

The Lagrangian of the two-flavor NJL model with scalar-isoscalar and
pseu\-do\-sca\-lar-isovector interactions is
\begin{equation}
{\cal L}=\bar q({\rm i}\partial^\mu \gamma_\mu - m)q + 
 {\frac{1}{2 a^2}}\left( (\bar q q)^2 + 
 (\bar q{\rm i}\gamma_5 \mbox{\boldmath $\tau$} q)^2\right) \; ,  
 \label{eq:lagr}
\end{equation}
where $q$ is the quark field, $m$ is the current quark mass, and 
$1/a^2$ is the coupling constant. It is convenient to apply the 
formalism of effective action \cite{ItzZub} to Lagrangian (\ref{eq:lagr}). 
Details of this procedure are given in Ref. \cite{NBCRG96}. Meson fields 
are introduced in the usual way (partial bosonization), with $\Phi = 
(\Phi_0, \mbox{\boldmath $\Phi$})$ related to the sigma and 
the pion mean field. At the quark-loop level the 
effective action is
\begin{equation}
{I}(\Phi) = \int d^4x \left ( \half{a^2} \Phi^2 
 - a^2 m \Phi_0 + \half{a^2} m^2 \right )
 - \half{\rm Tr}\,\ln (D^{\dagger }D)  \;,  
\label{eq:Seffq}
\end{equation}
where  $D$ is the Dirac operator , 
 $D = \partial_\tau - {\rm i}{\mbox{\boldmath $\alpha$}
  \cdot \mbox{\boldmath $\nabla$} } 
    + \beta \Phi_0 + {\rm i} \beta \gamma_5 
   {\mbox{\boldmath $\tau$}} \cdot {\mbox{\boldmath $\Phi$}}$.
We work in Euclidean space-time ($\tau$, ${\mbox{\boldmath $x$}}$). 
In Eq.~(\ref{eq:Seffq}) we have replaced the usual ${\rm Tr}\,\ln D$ term 
with $\half{\rm Tr}\,\ln (D^{\dagger }D)$, which is allowed in the 
absence of anomalies. In fact, this replacement is necessary for the
introduction of the proper-time regulator \cite{pt} used in many NJL 
calculations, and also in this paper.

Meson loops bring an additional term to the effective action 
\cite{NBCRG96,ItzZub}
\begin{equation}
{\Gamma}(\Phi) =  {I}(\Phi) + \half {\rm Tr}\,\ln ({K}^{-1})  \; . 
 \label{eq:Seffm}
\end{equation}
The inverse {\em meson propagator} matrix $K$ is defined as 
 $K^{-1}_{ab}(x,y) = \frac{\delta^2 I \left( \Phi \right)}
  {\delta \Phi_a(x) \delta \Phi_b(y)}$.
In Eqs.~(\ref{eq:Seffq},\ref{eq:Seffm}) 
${\rm Tr}$ denotes the full trace, including functional space, 
isospin, and in addition color and spinor trace for quarks.
In the $N_c$-counting scheme, the quark loop term ${I}(\Phi)$
is the leading contribution of order ${\cal O}(N_c)$, and the 
meson loop term $\frac{1}{2}{\rm Tr}\,\ln {K}$ is of order ${\cal O}(1)$.
Thus the one-meson-loop contributions give the first correction to
the leading-$N_c$ results.

\begin{figure}[b]
\xslide{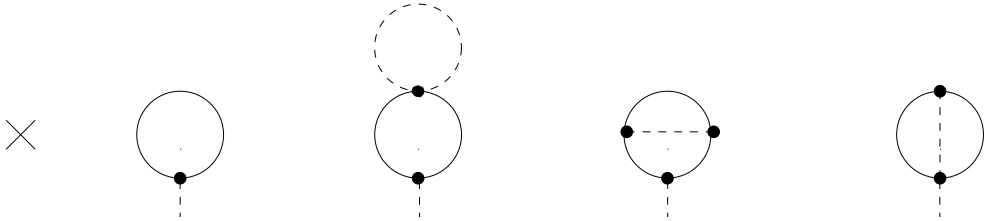}{3cm}{30}{370}{560}{490}
\caption{Diagramatic representation of Eq.~(\ref{gap0}). The cross
represents the first term, the one-quark-loop contribution 
corresponds to the second term, and meson-loop terms 
represent subsequent terms. The solid lines represent the 
quark propagator $1/(D^{\dagger} D)$, the dased lines correspond to 
the meson propagators $K$ of Eqs.~(\ref{prop}), the external 
dased line represents scalar-isoscalar coupling, and the vertices follow
from the form of $(D^{\dagger} D)$.} 
\label{fig:0}
\end{figure}

Using standard methods, Green's functions can be obtained from 
Eq.~(\ref{eq:Seffm}) via differentiation with respect to mean 
meson fields.
Of particular importance is the one-point function, which gives the
expectation value of the sigma field.  The condition
\begin{eqnarray}
\label{GAP} 
\frac{\delta \Gamma(\Phi)} {\delta \Phi_0(x)}_{\mid \Phi_0(x)=S} 
= && a^2 (S-m) - \half {\rm Tr}
\left( (D^{\dagger} D)^{-1} \frac{\delta (D^{\dagger }D)}{\delta \Phi_0(x)}
\right )_{\Phi_0(x)=S}  \nonumber \\ 
&& + \half {\rm Tr} \left ( K \frac{\delta K^{-1}}
{\delta \Phi_0(x)} \right)_{\Phi_0(x)=S} = 0
\end{eqnarray}
yields the equation for the vacuum expectation value of $\Phi_0$, 
which we denote by $S$. 
Introducing 
\begin{eqnarray}
\label{prop}
K_\sigma(S,Q^2) & = & \left ( 4 N_c f(S,Q^2)(Q^2+4S^2) + 
a^2 m/S \right )^{-1} \; , \nonumber \\
K_\pi(S,Q^2) & = & \left ( 4 N_c f(S,Q^2) Q^2 + a^2 m/S  \right )^{-1} \;,
\end{eqnarray}
and retaining terms up to order ${\cal O}(N_c^0)$, 
Eq.~(\ref{GAP}) can be written in the form \cite{NBCRG96}
\begin{eqnarray}
\label{gap0}
& & a^2 \left(S - m \right) - 8 N_c \, S g(S) \nonumber \\
& & + S \frac{N_c}{4 \pi^4} \int d^4 Q 
   \left\{ \left [2 f(S,0) + \frac{d}{dS^2} 
 \left (f(S,Q^2)(Q^2 + 4 S^2) \right ) \right]
   K_\sigma(S,Q^2) \right. \nonumber \\
& & + \left. 3 \left [2 f(S,0) + \frac{d}{dS^2} f(S,Q^2) Q^2 \right]  
 K_\pi(S,Q^2) \right\} = 0.
\end{eqnarray}
Functions $g$ and $f$ in the above expressions are the {\em quark 
bubble functions}. Their form is very simple if no cut-offs were present.
In this case we would have 
\mbox{$g(S) = \int {d^4k \over (2\pi)^4} {1 \over k^2 + S^2}$} and
\mbox{$f(S,Q^2) = \int {d^4k \over (2\pi)^4} {1 \over k^2 + S^2}
{1 \over (k+Q)^2 + S^2 }$}, and Eq.~(\ref{gap0}) could be interpreted
via standard Feynman diagrams (see Fig.~\ref{fig:0}).
In the presence of a cut-off these functions are complicated. 
In the case of the proper-time cut-off \cite{pt} used here
we have \cite{NBCRG96}
\begin{equation}
\label{g0r}
g(S) = \int {d^4k \over (2\pi)^4} \int\limits_{\Lambda_f^{-2}}^{\infty}
ds \, \exp\left\{-s [k^2+S^2]\right\} = {\Lambda_f^2 \over 16 \pi^2} 
\, E_2\left[{S^2 \over \Lambda_f^2} \right]
\end{equation}
and
\begin{eqnarray}
\label{f0r}
f(S,Q^2) &=& \int {d^4k \over (2\pi)^4} \int\limits_{\Lambda_f^{-2}}^{\infty}
ds \, s \int\limits_0^1 du \exp\left\{-s [k^2 + S^2 + u(1-u) Q^2] \right\} 
 \nonumber \\
&=& {1 \over 16 \pi^2} \int\limits_0^1 du \, E_1\left[{S^2 \over \Lambda_f^2}
+ u (1-u) {Q^2 \over \Lambda_f^2} \right],
\end{eqnarray}
where $\Lambda_f$ is the quark cut-off, and 
the exponential integral is defined as 
\mbox{$E_n(x) \equiv \int\limits_1^{\infty} dt \, {e^{-xt} \over t^n}$}.

The one meson-loop gap equation (\ref{gap0}) requires also the
introduction of a regulator for meson momenta. In other words,  
we have to regularize the divergent integral over $d^4Q$.
In Ref. \cite{NBCRG96} this was achieved by the substitution
\mbox{$\int d^4Q  \longrightarrow \pi^2 \int\limits_0^{\Lambda_b^2} 
 dQ^2 \, Q^2$}, where $\Lambda_b$ was the four-dimensional Euclidean meson
momentum cutoff. 
In the present study at finite temperatures, 
we employ the
three-dimensional cutoff procedure, i.e., we make the replacement
\begin{equation}
\label{3dc}
\int d^4Q  \longrightarrow 4 \pi \int d\omega 
\int\limits_0^{\Lambda_b} dq \, q^2 \,\,\,\,\,\, ,
\end{equation}
where $Q=(\omega,{\bf q})$ and $q = |{\bf q}|$. The form
(\ref{3dc}) is convenient for the implementation of the boundary
conditions satisfied by temperature Green's functions.

\vfill

\section{Finite temperature}
\label{sec:gap}

For calculations at finite temperature $T$ we shall adopt the imaginary time
formalism \cite{Kapusta}. 
This can be done by making the following replacement in the
quark momentum integrals
\begin{equation}
\label{itf}
\int {d^4k \over (2\pi)^4} F(k) = \int {dE \over 2\pi} \int
{d^3k \over (2\pi)^3} F(E,{\bf k}) \rightarrow T \sum_{j=-\infty}^{\infty}
\int {d^3k \over (2\pi)^3} F(E_j,{\bf k}).
\end{equation}
Here $F(k)=F(E,{\bf k})$ is an arbitrary integrand, 
and the sum runs over the fermionic 
Matsubara frequencies $E_j = (2j+1)\pi T$. The integral over the meson
four-momenta should be also replaced by the sum of the form (\ref{itf}).
In this case, however, the sum runs over the bosonic Matsubara
frequencies $\omega_n = 2\pi n T$.
With this prescription we can turn to
the calculation of the functions which are the finite temperature
analogs of $g(S)$ and  $f(S,Q^2)$.
We find
\begin{eqnarray}
g(S,T) &=& T \sum_j \int {d^3k \over (2\pi)^3} 
\int\limits_{\Lambda_f^{-2}}^{\infty} ds \, 
\exp\left\{ -s \left[ E_j^2
+ {\bf k}^2 + S^2 \right] \right\} \nonumber \\
&=& {T \Lambda_f \over 8 \pi^{{3 \over 2}} } \sum_j
E_{3 \over 2} \left[{{S^2+E_j^2} \over \Lambda_f^2} \right]
\end{eqnarray}
and
\begin{eqnarray}
\!\!\!\!\!\!\!\!\!\!\!\!\!\!\!& &f(S,n,q,T) = 
T \sum_j \int {d^3k \over (2\pi)^3} 
\int_{\Lambda_f^{-2}}^{\infty} ds \, s \int\limits_0^1 du \times 
\nonumber \\ 
\!\!\!\!\!\!\!\!\!\!\!\!& & \exp\left\{ -s \left[ 
S^2 + u(1-u)(\omega_n^2 + {\bf q}^2) + \left[ {\bf k}
- {\bf q} (1-u) \right]^2 + \left[E_j - \omega_n (1-u) \right]^2
\right] \right\}  \nonumber \\ 
\!\!\!\!\!\!\!\!\!\!\!\!& &=  {T \over 8 \pi^{{3 \over 2}} \Lambda_f} \sum_j
\int\limits_0^1 du \, E_{1\over 2}\left[{S^2 \over \Lambda_f^2} +
u(1-u) {\omega_n^2 + {\bf q}^2 \over \Lambda_f^2} \right. 
  +  \left. {[E_j-\omega_n(1-u)]^2 \over \Lambda_f^2} \right] \;.
\end{eqnarray}
Analogously, the inverse meson propagators become
\begin{eqnarray}
\label{propT}
 K_\sigma(S,n,q,T) & = & \left ( 
4 N_c f(S,n,q,T)(\omega_n^2 + {\bf q}^2+4S^2) + 
  a^2 m/S \right )^{-1} \; , \nonumber \\ 
 K_\pi(S,n,q,T) & = & \left 
 ( 4 N_c f(S,n,q,T)(\omega_n^2 + {\bf q}^2) + a^2 m/S  \right )^{-1} \;.
\end{eqnarray}
Finally, we can write the finite-temperature analog
of Eq.~(\ref{gap0}):
\begin{eqnarray}
\label{gapT}
& & a^2 \left(S - {m} \right) - 8 N_c S \, g(S,T) + 
 {2 S N_c T \over \pi^2} \sum_n \int\limits_0^{\Lambda_b} dq \, q^2 \times 
\nonumber \\
& & \left\{ \left [2 f(S,0,0,T) + \frac{d}{dS^2} \left (f(S,n,q,T)
(\omega_n^2 + {\bf q}^2 + 4S^2) \right ) \right]
 K_\sigma(S,n,q,T) \right. \nonumber \\
& & + \left. 3 \left [2 f(S,0,0,T) + \frac{d}{dS^2} f(S,n,q,T) 
(\omega_n^2 + {\bf q}^2) \right]  
 K_\pi(S,n,q,T) \right\} = 0 \;.\nonumber \\
\end{eqnarray}
If chiral symmetry is broken, then the above equation has a 
nontrivial solution for $S$. 
The quark condensate and $S$ are related by the formula
\begin{equation}
\label{qq}
\langle \overline{q}q \rangle = - a^2 (S - m) \;,
\end{equation}
which follows immediately from the fact that 
 $\langle \overline{q}q \rangle = \delta \Gamma(\Phi)/\delta m$
and Eq.~(\ref{eq:Seffm}).

\section{Low-temperature expansion in the chiral limit}
\label{sec:lowT}

Before presenting our numerical results for $\langle {\overline q} q
\rangle_T$
let us consider the low-temperature expansion. As shown by
Gasser and Leutwyler \cite{GL}, {\em in the chiral limit}
the low-temperature expansion of the quark condensate has the form
\begin{equation}
\label{eq:gl}
\langle \overline{q} q \rangle_T = \langle \overline{q} q \rangle_0 
\left ( 1 - \frac{T^2}{8 F_\pi^2} - \frac{T^4}{384 F_\pi^4} + ... \right ) .
\end{equation}
First, let us do the $N_c$ counting in this formula. Since
 $F_\pi \sim {\cal O}(\sqrt{N_c})$, subsequent terms in the expansion are 
suppressed by $1/N_c$. Since our one-meson-loop 
calculation accounts for first
subleading effects in the  $1/N_c$ expansion, we can hope for reproducing 
only the $T^2$ term in Eq.~(\ref{eq:gl}). Further terms would require
more loops.

Using standard techniques \cite{Kapusta}, 
the sum over the bosonic Matsubara
frequencies in Eq.~(\ref{gapT}) can be converted to a contour integral
in the complex energy plane. By deforming this contour
we collect all contributions from the singularities of the
integrand, weighted with the thermal Bose distribution. At low
temperatures, the dominant contribution comes from the lowest
lying pion pole, and other singularities are negligible. 
Thus, the third term in
(\ref{gapT}) becomes
\begin{equation}
\label{ae1}
{3T \over \pi^2} \sum_n \int\limits_0^{\Lambda_b} dq \, q^2
{1 \over \omega_n^2 + q^2} = {3 \over 2\pi^2 }
\int\limits_0^{\Lambda_b} dq \, q \left[ 1 + {2 \over e^{q/T} - 1} \right].
\end{equation}
Writing Eq. (\ref{ae1}) we have approximated the function 
$f(M,n,q,T)$, appearing in the pion propagator, by its value at $n=q=0$. 
For sufficiently large cutoff $\Lambda_b$, the integral over the thermal 
distribution function in (\ref{ae1}) can be expressed by the Riemann
zeta function $\zeta(2) = \pi^2/6$. Thus, the final result for
(\ref{ae1}) is $ 3\Lambda^2_b/4 \pi^2 + T^2/2$.
Inserting the above result into the gap equation (\ref{gapT}) we find,
with $m=0$, the following equality:
\begin{equation}
\label{ae2}
h(S,T) \equiv a^2 - 8N_c g(S,T) + {3\Lambda_b^2 \over
4 \pi^2} + \half T^2 = 0.
\end{equation}
Eq. (\ref{ae2}) defines implicitly the function $S(T)$,
which satisfies the equation
\begin{equation}
\label{ae3}
{dS \over dT^2} = - 
{ \partial h(S,T) / \partial T^2 \over \partial h(S,T)   /  \partial S}
= \left[ 16 N_c
{\partial g(S,T) \over \partial S} \right]^{-1}.
\end{equation}
Here we have neglected the term $\partial g(S,T) / 
\partial T^2$, since it is exponentially suppressed by the factor
 $\exp(-S/T)$. Furthermore, the leading-$N_c$ term on the 
right hand side of (\ref{ae3}) can be rewritten using the 
relations \cite{NBCRG96}
 $\partial g(S,T)/ \partial S = -2S f(S,0)$ and 
 \mbox{$4 N_c f(S,0) = \overline{F}_{\pi}^2/S^2$}, 
 where $\overline{F}_{\pi}$ is the leading-$N_c$ 
piece of the pion decay constant. 
Collecting these equalities we 
arrive at $dS/dT^2 = -S/(8\overline{F}_{\pi}^2)$, which finally gives
\begin{equation}
\label{ae4}
S(T) = S(0) \left[ 1 - {T^2 \over 8\overline{F}_{\pi}^2} \right].
\end{equation}
Proportionality (\ref{qq}) implies that the above expression
coincides (in the large $N_c$ limit) with Eq.~(\ref{eq:gl}).
Hence our method is consistent with a basic requirement of chiral
symmetry at the one-meson-loop level.

\section{Results}
\label{sec:res}

In the exact chiral limit the 
model has 3 parameters: $a$, $\Lambda_f$, and $\Lambda_b$. 
In this paper we fix arbitrarily $\Lambda_b/\Lambda_f = \half$.
The remaining
2 parameters are fixed by reproducing the physical value of
 $F_\pi=93{\rm MeV}$ and 
a chosen value for $\langle \overline{q} q \rangle_0$. 
For the case of $m \neq 0$ we have an extra parameter, $m$, which is fitted
by requiring that the pion has its physical mass.
We compare results with meson loops to results with the quark loop
only ($\Lambda_b = 0$).  Parameters for the two calculations
are adjusted in such a way, that the values of $F_\pi$, 
 $\langle \overline{q} q \rangle_0$, and $m_\pi$ are the same.

The calculation of $F_\pi$ with meson loops, although 
straightforward, is rather tedious, so we 
do not present it here. The method has been presented in detail in 
Ref.~\cite{NBCRG96,thesis}. The only difference in our calculation 
is that the three-momentum cut-off (\ref{3dc}) rather than
the four-momentum cut-off of Ref.~\cite{NBCRG96} is used.

\begin{figure}[b]
\xslide{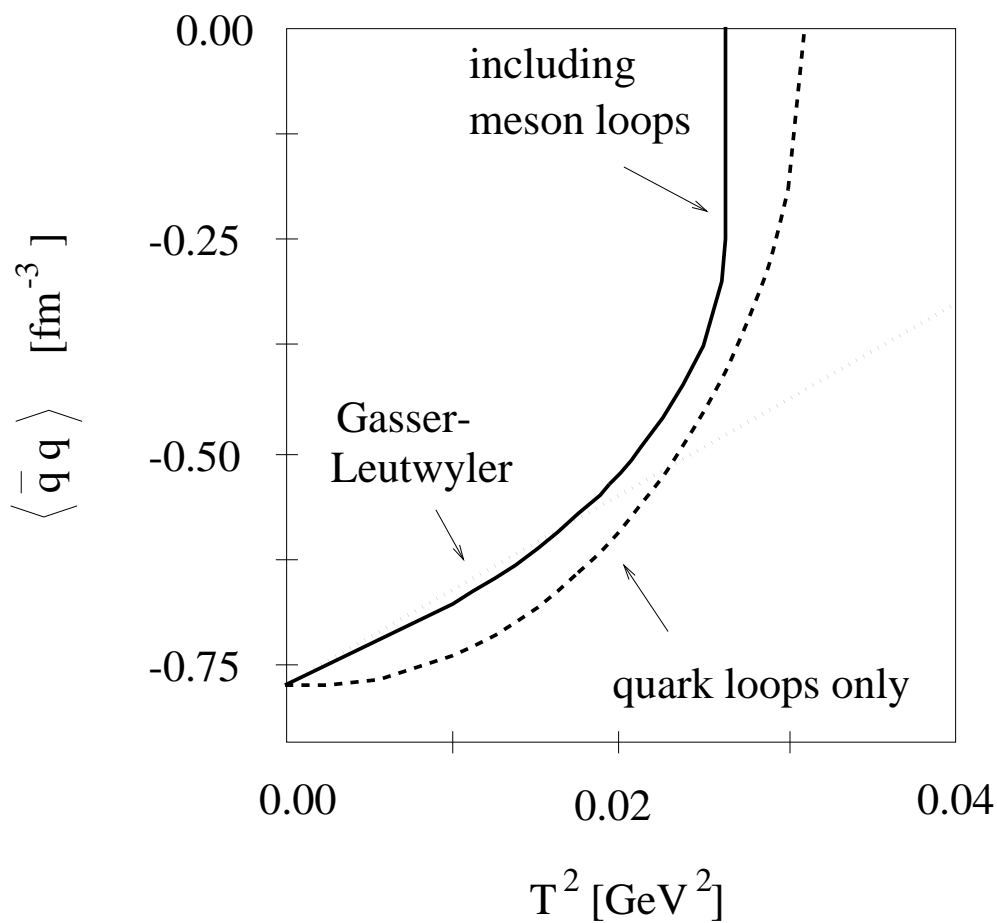}{13.5cm}{45}{160}{550}{680}
\caption{Dependence of of the quark condensate on $T^2$ 
in the chiral limit $m_\pi=0$. The curves correspond to 
the calculation
with meson loops (solid line), with quark loops only (dashed line),
and the lowest-order chiral expansion (dotted line). The parameters
for the solid line and dashed line are adjusted in such a way that
$F_\pi = 93{\rm MeV}$ and $\langle \overline{q} q \rangle_0 
 =-(184{\rm MeV})^3$. For the solid line $a=175{\rm MeV}$, 
 $\Lambda_f = 723{\rm MeV}$, and $\Lambda_b = \half \Lambda_f$, 
whereas for the
dashed line $a=201{\rm MeV}$, $\Lambda_f = 682{\rm MeV}$, and
 $\Lambda_b = 0$. }
\label{fig:1}
\end{figure}
Figure \ref{fig:1} shows the dependence of $\langle \overline{q} q \rangle$
on $T^2$. The solid line represents the case with meson loops. We note that
at low temperatures the curve has a finite slope, as requested by
Eq.~(\ref{ae4}). The slope is close to the leading-order 
Gasser-Leutwyler result (dotted curve). As explained earlier, 
the slopes would overlap in the large-$N_c$ limit. 
This behavior is radically different from the case with quark loops 
only (dashed curve). In this case at low temperatures 
 \mbox{$\langle \overline{q} q \rangle_T - 
  \langle \overline{q} q \rangle_0 \sim e^{-M/T}$}, 
where $M$ is the 
mass of the constituent quark. All derivatives of this function vanish
at $T=0$, and $\langle \overline{q} q \rangle$ is flat at the origin.
We can also see from the figure that the fall-off of the condensate
is faster when the meson loops are included. 
In fact, for the parameters of Fig.~\ref{fig:1} we have an interesting
phenomenon. At $T = 162{\rm MeV}$ the condensate abruptly jumps to 0. There is 
a first-order phase transition, with a latent heat necessary to melt
the quark condensate. Such a behavior is not present 
in the case of calculations without meson loops \cite{HK85,BMZ87,Lutz92}.
We note that with meson loops present the chiral restoration 
temperature is $162{\rm MeV}$, {\em i.e.} about 10\% 
less than $176{\rm MeV}$ of the quark-loop-only case.
\begin{figure}[b]
\xslide{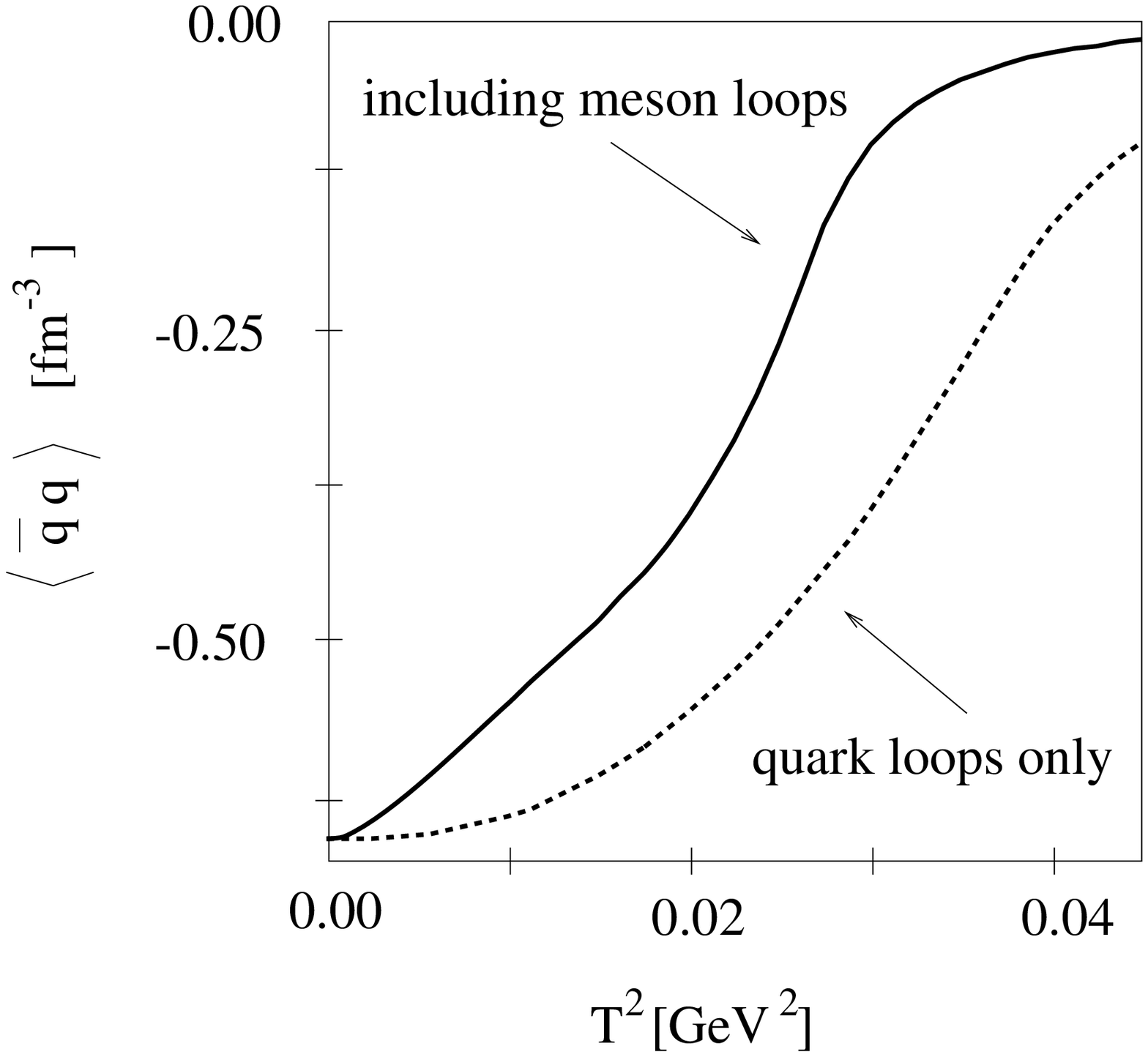}{13.5cm}{45}{160}{550}{680}
\caption{Same as Fig.~\ref{fig:1} for $m_\pi=139~{\rm MeV}$, 
 $F_\pi = 93{\rm MeV}$, and $\langle \overline{q} q \rangle_0 
 =-(174{\rm MeV})^3$.
For the solid line $a=164{\rm MeV}$, 
 $\Lambda_f = 678{\rm MeV}$, $\Lambda_b = \half \Lambda_f$, and 
 $m = 15{\rm MeV}$, whereas for the
dashed line $a=175{\rm MeV}$, 
 $\Lambda_f = 645{\rm MeV}$, $\Lambda_b = 0$, and $m = 15{\rm MeV}$. }
\label{fig:2}
\end{figure}

Figure \ref{fig:2} shows the same study, but for the physical
value of $m_\pi$. We note that now $\langle \overline{q} q \rangle$ 
(solid line) is also flat at the origin, since the pion is
no more massless, and at low $T$ we have 
 \mbox{$\langle \overline{q} q \rangle_T - \langle \overline{q} q \rangle_0
 \sim e^{-m_\pi/T}$}.
Nevertheless, the region of this flatness is small, and at 
intermediate temperatures the curve remains close to the
Gasser-Leutwyler expansion. We note again that meson loops 
considerably speed up
the melting of the condensate compared to the case of quark loops only.
However, there is no first-order phase transition such as in 
Fig.~\ref{fig:1}. Instead, we observe a smooth cross-over typical for the case
of $m \neq 0$. 

The faster change of the quark condensate in our study
is not surprising.
It is caused by the presence of light pions which 
are known to play a dominant role
at low-temperatures \cite{Heid}. 
The behavior of  $\langle \overline{q} q \rangle$ 
reflects this general feature.
Concluding, we stress that the inclusion of meson loops 
in the NJL model 
qualitatively and quantitatively changes the results 
in comparison to the 
calculations at the quark-loop level. In particular, we find finite slope of
 $\langle \overline{q} q \rangle$ vs. $T^2$ at the origin in the chiral
limit, faster melting of the condensate, and lower chiral restoration 
temperature. 

%\bibliographystyle{npsty}
%\bibliographystyle{npa96}
%\bibliography{wf}

\end{document}